\documentclass[prb,reprint,twocolumn,superscriptaddress,noshowpacs,notitlepage,longbibliography,10pt,citeautoscript]{revtex4-2}%

\usepackage{graphicx,bm,times}
\graphicspath{ {./figures/main/} }
\usepackage{amsmath}
\usepackage{amsfonts}
\usepackage{amssymb}
\usepackage{mathtools}
\usepackage{color}
\usepackage{hyperref}
\hypersetup{
	colorlinks = true,
	allcolors = {blue}
}

\begin{document}

\title{Dichotomy of electron-phonon interactions in the delafossite PdCoO$_2$: From weak bulk to polaronic surface coupling}

\author{Gesa-R. Siemann}
\affiliation{SUPA, School of Physics and Astronomy, University of St Andrews, St Andrews KY16 9SS, UK}

\author{Philip A. E. Murgatroyd}
\affiliation{SUPA, School of Physics and Astronomy, University of St Andrews, St Andrews KY16 9SS, UK}

\author{Tommaso Antonelli}
\affiliation{SUPA, School of Physics and Astronomy, University of St Andrews, St Andrews KY16 9SS, UK}

\author{Edgar Abarca Morales}
\affiliation{SUPA, School of Physics and Astronomy, University of St Andrews, St Andrews KY16 9SS, UK}
\affiliation{Max Planck Institute for Chemical Physics of Solids, 01187 Dresden, Germany}

\author{Seunghyun Khim}
\affiliation{Max Planck Institute for Chemical Physics of Solids, 01187 Dresden, Germany}

\author{Helge Rosner}
\affiliation{Max Planck Institute for Chemical Physics of Solids, 01187 Dresden, Germany}

\author{Matthew D. Watson}
\affiliation{Diamond Light Source Ltd., Harwell Science and Innovation Campus, Didcot OX11 0DE, UK}

\author{Andrew P. Mackenzie}
\affiliation{Max Planck Institute for Chemical Physics of Solids, 01187 Dresden, Germany}
\affiliation{SUPA, School of Physics and Astronomy, University of St Andrews, St Andrews KY16 9SS, UK}

\author{Phil D. C. King}
\email{pdk6@st-andrews.ac.uk}
\affiliation{SUPA, School of Physics and Astronomy, University of St Andrews, St Andrews KY16 9SS, UK}

\date{\today}

\begin{abstract} 
The metallic delafossites host ultra-high mobility carriers in the bulk, while at their polar surfaces, intrinsic electronic reconstructions stabilise markedly distinct electronic phases, from charge-disproportionated insulators, to Rashba-split heavy-hole gases and ferromagnetic metals. The understanding of these phases has been strongly informed by surface spectroscopic measurements, but previous studies have been complicated by the presence of spatially varying terminations of the material surface. Here, we demonstrate the potential of microscopic-area angle-resolved photoemission to overcome these challenges. Our measurements of the model compound PdCoO$_2$ yield extremely high-quality spectra of the electronic structure, which allows us to place new experimental constraints on the weak electron-phonon coupling in the bulk of PdCoO$_2$, while revealing much stronger interactions at its surfaces. While the CoO$_2$-terminated surface exhibits a conventional weak-coupling behavior, our measurements reveal surprising spectroscopic signatures of polaron formation at the Pd-terminated surface, despite its pronounced metallicity. Together, our findings reveal how mode and symmetry-selective couplings can markedly tune the electron-phonon interactions in a single host material, here opening routes to stabilise surprisingly-persistent polaronic quasiparticles.

\end{abstract}

\maketitle

\section{Introduction}
The coupling of electrons to the collective vibrations of the crystal lattice plays a crucial role in dictating a material's electrical and thermal properties, as well as underpinning the emergence of a host of collective states including superconductivity \cite{migdal1958interaction, eliashberg1960interactions,stronginEnhancedSuperconductivityLayered1968,mcmillanTransitionTemperatureStrongCoupled1968a,lanzaraEvidenceUbiquitousStrong2001}, charge density waves \cite{peierlsZurTheorieElektrischen1930,luoElectronicNatureCharge2022,xieElectronphononCouplingCharge2022} and the formation of polarons; quasiparticle excitations where the electrons or holes become dressed by the lattice vibrations \cite{dykmanRootsPolaronTheory2015,franchiniPolaronsMaterials2021}. Polarons can, in turn, lead to the formation of in-gap states, modifying optical properties, electrical conductivity, and chemical reactivity of materials, while they are also thought to play a central role in governing a diverse array of emergent materials phenomena such as colossal magneto resistance \cite{franchiniPolaronsMaterials2021, alexandrovAdvancesPolaronPhysics2010, bredasPolaronsBipolaronsSolitons1985}. Meanwhile, minimising coupling to the lattice is crucial, for example, to realise ultra-high electrical conductivities - of key importance to low-energy electronics - and to optimise thermoelectrics. Understanding the conditions under which electron-phonon coupling strengths can be effectively tuned is, therefore, of key importance.

We take the metallic delafossite PdCoO$_2$ as a model compound in which to probe these effects. PdCoO$_2$ is a layered material, and can be thought of as a natural heterostructure of Pd$^{1+}$ and $($CoO$_2)^{1-}$ layers (Fig.~\ref{fig:spatial}a))~\cite{hicksQuantumOscillationsHigh2012a,kushwahaNearlyFreeElectrons2015a,mackenziePropertiesUltrapureDelafossite2017a}. The bulk Pd layers are highly conducting, while the bulk CoO$_2$ layers are band insulating~\cite{Shannon1971,mackenziePropertiesUltrapureDelafossite2017a,kushwahaNearlyFreeElectrons2015a}. At the surface, however, the nominal net charge transfer of one electron from the Pd layer to the surrounding CoO$_2$ layers is disrupted, leading to a significant hole (electron) doping of its exposed CoO$_2$ (Pd) terminating surface layer with respect to the bulk charge count~\cite{kimFermiSurfaceSurface2009,nohAnisotropicElectricConductivity2009a,sunkoMaximalRashbalikeSpin2017b, yimQuasiparticleInterferenceQuantum2021, mazzolaItinerantFerromagnetismPdterminated2018a, yimAvoidedMetallicityHoledoped2023} (Fig.~\ref{fig:spatial}a)), leading to metallic states at both surfaces. 

Here, we utilise microscale spatially- and angle-resolved photoemission spectroscopy ($\mu$-ARPES) to isolate contributions to the electronic structure from these distinct surface terminations. The extremely high quality data obtained allows us to perform a detailed investigation of their quasiparticle dynamics. These measurements indicate a very weak electron-phonon coupling in the bulk, but we find that this becomes strongly enhanced at the surface. For the CoO$_2$ termination, we show that the interactions are nonetheless well described within a conventional weak-coupling Migdal-Eliashberg framework. In contrast, for the Pd-terminated surface, we observe spectroscopic signatures of polaron formation, despite the highly itinerant electron gas that exists at this surface. We attribute the stabilisation and persistence of such a polaronic regime to the weak screening of the out-of-plane Pd-O stretch mode, and further demonstrate its extreme sensitivity to surface adsorption, opening new routes for tuning the strength of polaronic coupling in metallic 2D electron systems.

\section{Spectro-microscopy of P\lowercase{d}C\lowercase{o}O$_2$}

\begin{figure*}[t]
    \centering
    \includegraphics[width=\textwidth]{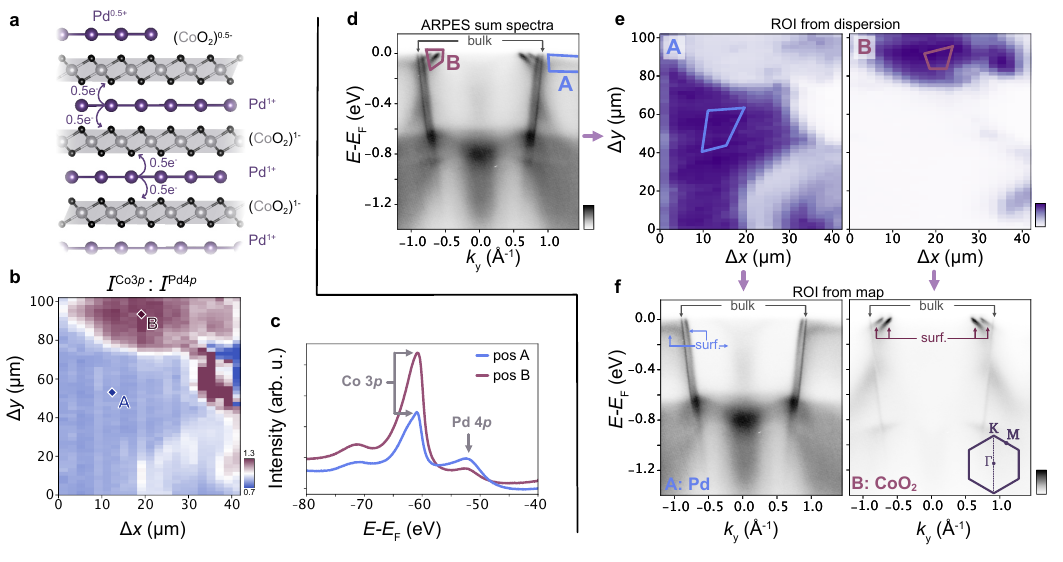}
    \caption{(a) Schematic of the layered crystal structure of PdCoO$_2$, showing the nominal interlayer charge transfer between Pd$^{1+}$ and (CoO$_2$)$^{1-}$ layers in the bulk, and the resulting doping of $\approx0.5$ holes/Co-atom (electrons/Pd-atom) at the surface. (b) Spatial variations in the intensity ratio of the measured Co 3$p$ ($I^{\text{Co~3}p}$) and Pd 4$p$ ($I^{\text{Pd~4}p}$) core level intensities, measured using a photon energy of $h\nu = 170$~eV. (c) Representative core level spectra for two different regions of the sample ('A' and 'B', as indicated by the diamond markers in (b)). (d) Valence band spectrum measured along the $\Gamma$-K direction ($h\nu=90$~eV, LH-pol., $T\approx\!35$~K), extracted from spatially-resolved ARPES data but integrated over the spatial region shown in (b). The formation of two distinct sets of surface states with hole-like ('B', purple region) and rather flat ('A', blue region) dispersions can be observed. (e) Selective integration over the 'A' and 'B' regions of interest (ROI) demonstrates a spatial confinement of these states to distinct regions of the sample, which are in good agreement with the contrast in our XPS mapping (b), indicating that they derive from Pd- and CoO$_2$-terminated regions of the sample, respectively. (f) A similar ROI analysis of the spatial mapping data confirms the presence of the hole-like states (labelled 'B' in (d)) on the CoO$_2$ termination, while the flat states are characteristic of the Pd-terminated surface.
   }
    \label{fig:spatial}
\end{figure*}

We show our $\mu$-ARPES measurements of PdCoO$_2$ in Fig.~\ref{fig:spatial}. The natural cleavage plane of the sample is located between the Pd and O layers. As a result, cleaving the sample will expose patches of both Pd- and CoO$_2$ terminated regions, randomly distributed across the surface (Fig.~\ref{fig:spatial}(a)). These regions have a spatial extent of only a few 10's of $\mu$m's, which is smaller than the probing region of a typical ARPES light spot (often $\sim\!50-100$~$\mu$m), resulting in contributions of both surfaces typically being present in the measured electronic structures~\cite{mazzolaItinerantFerromagnetismPdterminated2018a, mazzolaTuneableElectronMagnon2022a}. Utilising capillary-focusing optics (see Appendix for Methods), here we achieve sufficient spatial resolution necessary to deterministically probe from a single terminating surface, while maintaining high enough spectral resolution to enable performing a detailed self-energy analysis as shown below.

Fig.~\ref{fig:spatial}(b) shows the spatial variation of the measured ratio of the Co 3$p\;$:$\;$Pd 4$p$ core-level spectral weight. Due to the surface sensitivity of photoemission spectroscopy, this spectral weight ratio -- in which we find marked variations across the probed sample region -- gives an effective metric of the terminating atomic species at the surface. Sharp boundaries are evident between regions of distinct contrast, suggesting that well-defined regions of Pd and CoO$_2$ surface termination are formed at the cleaved surface, with a spatial extent of several 10's of microns. We show in Fig.~\ref{fig:spatial}(c) high-resolution core-level spectra measured at two locations within regions of high and low relative Co:Pd spectral weight. In addition to their relative intensity variations, these also show subtle shifts in the binding energies of the Co $3p$ core level components from the different surface regions. From fitting this across our spatial mapping data (Supplementary Fig.~1), we find that these core level shifts correlate well with the identified surface terminations, showing a shift towards lower binding energies for the CoO$_2$-terminated surface with respect to the Pd-terminated one. This directly indicates a hole vs.\ electron doping for these surfaces~\cite{yimAvoidedMetallicityHoledoped2023, siemannSpinorbitCoupledSpinpolarised2023}, fully in line with the simple expectations for self-doping at a polar surface termination as outlined above.

Measuring an ARPES dispersion at each position of the same spatial mapping region, we can correlate spatial variations in surface termination with marked changes in the electronic band structure (Fig.~\ref{fig:spatial}(d-f)). Integrating our $\mu$-ARPES measurements over the full spatial region probed in Fig.~\ref{fig:spatial}(b) [comparable to the size of a typical synchrotron-ARPES probe beam], we find a complex near-$E_\mathrm{F}$ electronic structure (Fig.~\ref{fig:spatial}(d)), comprising both the known single metallic bulk band~\cite{kushwahaNearlyFreeElectrons2015a,nohAnisotropicElectricConductivity2009a} as well as multiple sets of surface states. However, integrating the intensity from our spatially-resolved ARPES measurements over regions of interest (ROIs, see Fig.~\ref{fig:spatial}(d)) in energy and momentum where two of these surface states have significant spectral weight reveals a marked spatial contrast. Taking the ROI over a nearly dispersion-less state near the Brillouin zone boundary (region A in Fig.~\ref{fig:spatial}(d)) yields significant spectral weight in the lower left portion of our spatial map, while the spectral weight from region B -- a set of hole-like states crossing the Fermi-level at a momentum of $k_\text{F} \approx -0.6$~\AA$^{-1}$ -- is mostly located in the upper portion of our probed spatial region (Fig.~\ref{fig:spatial}(e)). In fact, we find here an essentially one to one correspondence with the integrated spectral weight from this ROI analysis and the corresponding spatial variations in our core-level mapping (Fig.~\ref{fig:spatial}(b,e)). This allows us to conclusively assign the origin of these distinct surface states as deriving from the Pd- and CoO$_2$-terminated surface. 

While this is consistent with prior classifications~\cite{nohAnisotropicElectricConductivity2009a,kimFermiSurfaceSurface2009, sunkoMaximalRashbalikeSpin2017b,mazzolaItinerantFerromagnetismPdterminated2018a}, our data from spatial regions with defined single surface terminations (see ROI boxes in Fig.~\ref{fig:spatial}(e)) yield much sharper spectra than our typically obtained in the literature (Fig.~\ref{fig:spatial}(f)). They furthermore significantly simplify the interpretation of the data, devoid of the signatures of multiple surface terminations in a single measurement (cf. Fig.~\ref{fig:spatial}(d)) that have significantly hindered previous studies, particularly for the Pd-terminated surface \cite{mazzolaItinerantFerromagnetismPdterminated2018a,mazzolaTuneableElectronMagnon2022a}. The data quality obtained here, in turn, opens the possibility to perform more quantitative analysis of the spectra. In particular, we show in the following that it is possible to utilise these measurements to extract markedly distinct quasiparticle dynamics of the bulk and the surface states associated with each distinct surface termination.

\section{Weak bulk electron-phonon coupling}
\begin{figure}
    \centering
    \includegraphics[width=\columnwidth]{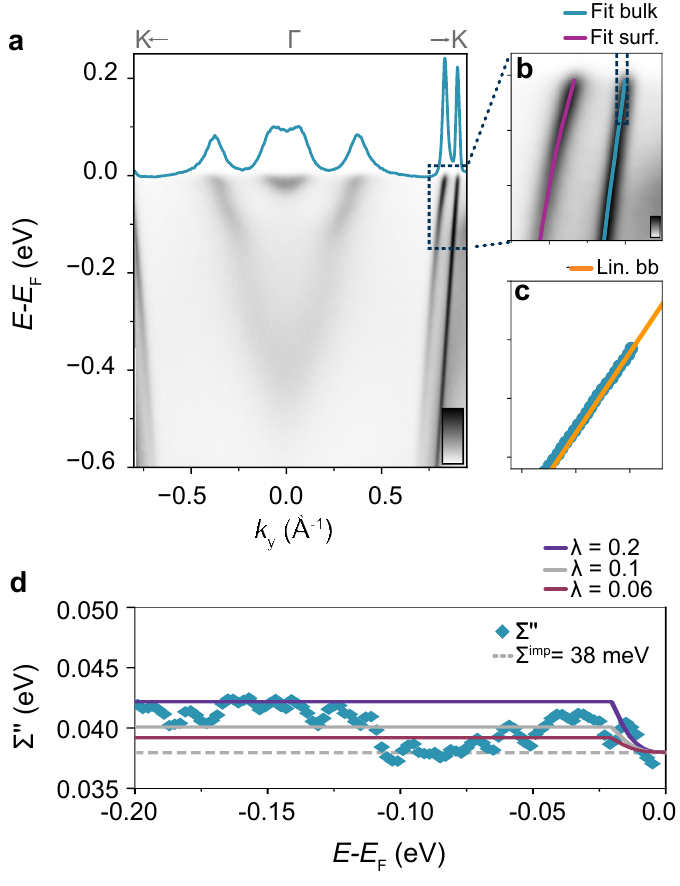}
    \caption{Weak electron-phonon coupling of bulk states in PdCoO$_2$. (a) Measured electronic structure ($h\nu=41$~eV) along the $\Gamma$-K direction from a Pd-terminated sample region. Measurements from this surface and at this photon energy show high spectral weight for the bulk state (the steep state with the highest $k_\mathrm{F}$ here). (b) shows a magnified view of the bulk states and a shifted surface copy of this, together with the peak positions from fits to MDCs. (c) Linear bare band approximation (orange solid line) fixed to the experimentally determined $k_{\text{F}}$. (d) Resulting imaginary component of the self-energy extracted from the fits to our measured dispersion for the bulk state using the bare band shown in (c). Migdal-Eliashberg simulations are also shown for electron-phonon coupling with varying coupling strengths, to a Debye phonon mode at an energy of 20~meV.
    }
    \label{fig:bulk}
\end{figure}
We first consider the electron-phonon coupling of the bulk states, which is predicted to be extremely weak~\cite{yaoOriginHighElectrical2024}. Fig.~\ref{fig:bulk}(a) shows measurements of the dispersion from a Pd-terminated sample region, but using a photon energy which enhances the spectral weight of the bulk states, to permit their detailed analysis. Two steeply dispersive states are observed crossing $E_\text{F}$ towards the K point. The one with the largest $k_\mathrm{F}$ is the bulk band, which is predominantly derived from the Pd orbitals~\cite{kushwahaNearlyFreeElectrons2015a,nohAnisotropicElectricConductivity2009a, usuiHiddenKagomelatticePicture2019}. A shifted copy of this state is observed with a lower $k_\mathrm{F}$, which results from the Pd-terminated surface, which we will return to below.

Fits to the dispersion of the bulk band extracted from momentum distribution curves (MDCs, Fig.~\ref{fig:bulk}(b)) yield a Fermi velocity of $8.1\pm0.2\times 10^{5}$~m/s along the $\Gamma-$K direction, consistent with previous  studies~\cite{hicksQuantumOscillationsHigh2012a, nohAnisotropicElectricConductivity2009a, hicksQuantumOscillationsMagnetic2015} of PdCoO$_2$. Recent calculations have predicted that this high Fermi velocity in part reflects a very weak electron-phonon coupling for the bulk states, which is caused by the low density of states at the Fermi level as well as geometric effects of the Fermi surface. In fact, these first principles calculations predict an electron-phonon coupling constant of only $\lambda=0.06$~\cite{yaoOriginHighElectrical2024}, consistent with very weak contributions of electron-phonon interactions to transport measurements~\cite{hicksQuantumOscillationsHigh2012a}. Our clearly-resolved measurement of the bulk state provides an opportunity to assess this weak coupling spectroscopically.  

\begin{figure*}
    \centering
    \includegraphics[width=\textwidth]{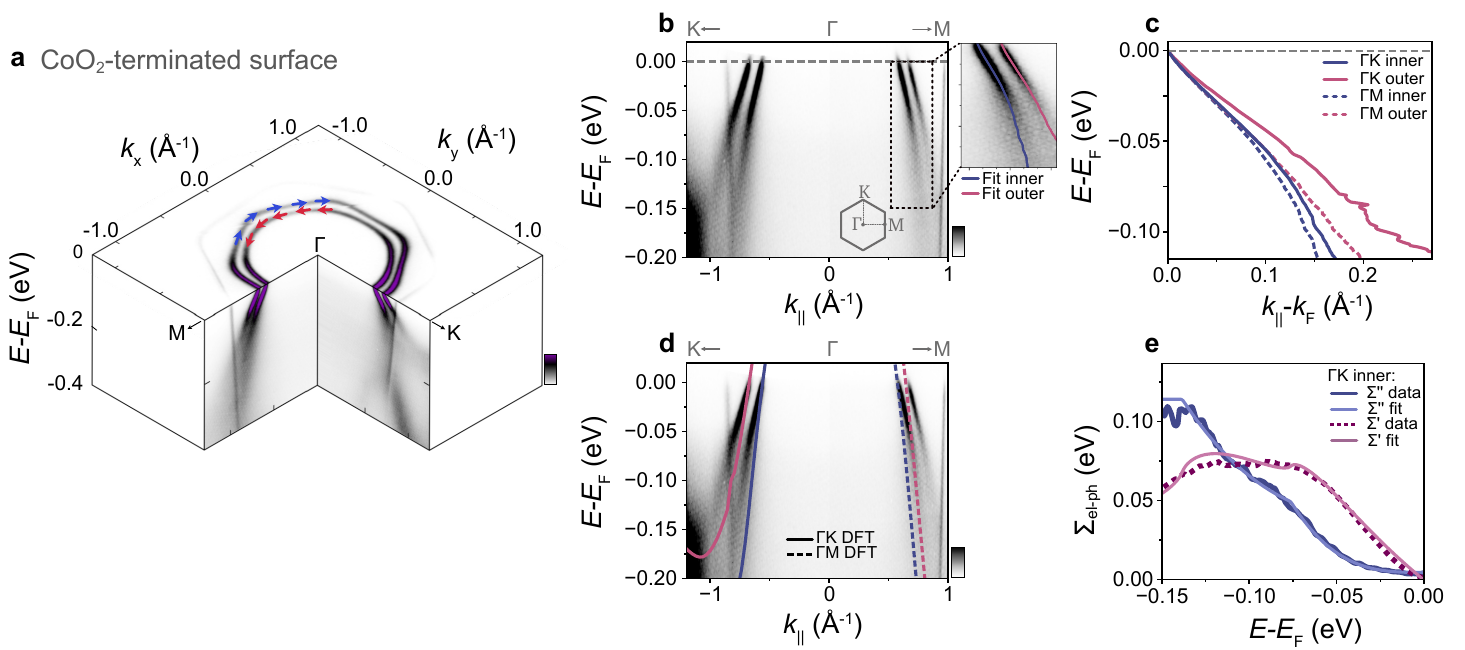}
    \caption{(a) Overview of the measured Fermi surface ($h\nu=110$~eV) of the CoO$_2$-terminated surface of PdCoO$_2$, and corresponding dispersions measured along the high symmetry $\Gamma$-M and $\Gamma$-K directions. The spin-polarised nature of these states is indicated by the blue and red arrows. (b) Measured dispersions ($h\nu=54$~eV) in the vicinity of the Fermi level, with fits to MDCs shown in the inset. (c) Peak positions from MDC fits along both directions, shifted to align their Fermi momenta. (d) Experimental data reproduced from (b), with results from DFT surface supercell calculations overlaid. The DFT calculations have been shifted by $92\pm 5$~meV and scaled in energy by a factor of 1.73 to align the Fermi crossings and band bottom along $\Gamma$-K, respectively (see Supplementary Fig.~3) and are used here as a bare band approximation. \textbf{e}.\ Experimentally-determined real and imaginary part of the self-energy ($\Sigma'$, purple dashed lines, and $\Sigma"$, navy blue lines, respectively) of the inner band along $\Gamma$-K.The Migdal-Eliashberg model including two oscillators is fitted to the imaginary part of the data (solid light blue line). The real part of the model self-energy is then found via a Kramers-Kronig transformation (pink solid line).
    }
    \label{fig:CoO}
\end{figure*}

In ARPES spectra, electron-phonon coupling is typically visible via a kink in the measured dispersion at the energy of characteristic phonon modes~\cite{hofmannElectronPhononCoupling2009c, hellsingElectronphononCouplingMetal2002a, kirkegaardSelfenergyDeterminationElectron2005c}. The absence of a clear kink-like feature in the dispersion here therefore already points to a weak electron-phonon coupling for the bulk states. Nonetheless, by fitting the dispersion, we can resolve a weak but finite enhancement of the effective mass in the vicinity of the Fermi level. Approximating the non-interacting bare band dispersion through a linear approximation (see Supplementary Fig.~2 for details), we estimate that the measured Fermi velocity is enhanced by approximately 4~\% (Fig.~\ref{fig:bulk}(c)), corresponding to $\lambda = 0.04 \pm 0.01 $. Spectroscopically resolving such a weak coupling in a steep band is challenging, and is sensitive to details of the bare band used (we find a value of $\lambda = 0.08 \pm 0.01 $ when using a parabolic bare band, see Supplementary Fig.~2). Nonetheless, this conclusively points to a weak interaction strength in the bulk electronic structure of PdCoO$_2$. We can further validate this via analysis of the measured linewidth. We show such an analysis in Fig.~\ref{fig:bulk}(d) (see also Supplementary Fig.~2b for the equivalent analysis using a parabolic bare band approximation). To estimate an upper limit on the electron-phonon coupling strength, and motivated by calculations of the Eliashberg spectral function shown in Ref.~\cite{yaoOriginHighElectrical2024}, we compare the imaginary component of the self-energy extracted from fits of our measured data to self-energy calculations for electron-phonon coupling to the known bulk phonon mode at $\omega=20$~meV, offset by an impurity scattering term ($\Sigma^{imp}=38$~meV). Despite the highly resolved spectra here, we cannot resolve any systematic signatures of electron-phonon interactions in the measured linewidth, which we can therefore constrain from this analysis to $\lambda\eqslantless0.1$ given the experimental uncertainty. This is entirely consistent with our estimates of an electron-phonon coupling strength of $\lambda=0.06\pm0.03$ from the above analysis of the real part of the self-energy using different bare band approximations (Fig.~\ref{fig:bulk}(c) and Supplementary Fig.~2). Together, this therefore points to a really rather weak electron-phonon interaction in the bulk, in good agreement with first-principles calculations for PdCoO$_2$~\cite{yaoOriginHighElectrical2024}, and even lower than that observed in highly-conductive elemental Cu ($\lambda=0.13$~\cite{allenEmpiricalElectronphononValues1987}). Combined with a high crystalline quality~\cite{sunkoPRX}, the extremely weak coupling between the carriers and the lattice can therefore be considered as a key contributor to the exceptionally high in-plane conductivity of PdCoO$_2$~\cite{yaoOriginHighElectrical2024, mackenziePropertiesUltrapureDelafossite2017a}. We show below that the situation is markedly different at the surface.

\section{Electron-phonon coupling at the C\lowercase{o}O$_2$-terminated surface}
Fig.~\ref{fig:CoO} shows our high-resolution $\mu$-ARPES data from a CoO$_2$-terminated surface region. The fast bulk band is still visible, forming the largest hexagonal Fermi surface in Fig.~\ref{fig:CoO}(a). Now, however, the spectral weight is dominated by a pair of hole-like bands which -- from our spectro-microscopy analysis above -- we can uniquely assign to the CoO$_2$-terminated surface. These have been shown to be a spin-split pair, hosting a surprisingly-large Rashba-like interaction~\cite{sunkoMaximalRashbalikeSpin2017b}, and forming a near-circular and hexagonal Fermi surface (Fig.~\ref{fig:CoO}(a)). Unlike for the bulk states, the measured dispersion of these surface states (Fig.~\ref{fig:CoO}(b)) exhibit clear spectroscopic signatures of electron-phonon interactions, with the formation of a pronounced kink in the dispersion accompanied by a stark increase in the measured linewidth with increasing binding energy. To quantify this, we have again fitted MDCs extracted from these dispersions (see, e.g., inset of Fig.~\ref{fig:CoO}(b)), benefiting from the substantially reduced linewidths we obtain here as compared to previous measurements~\cite{sunkoMaximalRashbalikeSpin2017b, nohAnisotropicElectricConductivity2009a}.

We show the extracted dispersions for the inner and outer bands along the high-symmetry directions, plotted relative to their respective Fermi wave vectors, $k_\text{F}$, in Fig.~\ref{fig:CoO}(c). The slopes vary in the vicinity of the Fermi level, largely due to the pronounced Fermi surface warping of the outer band, but all are rather shallow: the obtained effective masses are on the order of 7-12~$m_e$, nearly a factor of 10 higher than the value of $\approx\!1.3$~m$_e$ for the bulk states. Furthermore, all show the formation of a characteristic kink feature at a binding energy of $E-E_\text{F}\approx 75$~meV. We obtain the corresponding electronic self-energy from the fitted positions and linewidths, taking the form of the bare band dispersion from surface-projected density-functional theory calculations (see Appendix and Supplementary Figs. 3 and 4 for details). We note that this treatment works well for three of the bands, while $k_\mathrm{F}$ of the outer band along the $\Gamma$-M direction is underestimated. This reflects a systematic underestimation of the Rashba-type spin splitting along this direction in our calculations, and we thus focus in the following on the three bands where $k_\mathrm{F}$ is in good agreement between the calculations and experiment.

Using bare bands determined in this way, we obtain an average electron-boson coupling strength determined from the renormalisation of the extracted dispersion at the Fermi level $\lambda = v_F^{DFT}/v_F^{exp}-1=0.9\pm0.1$ (see Supplementary Fig.~5 for details of the analysis). This points to a rather strong interaction strength, with the total mass enhancement from many-body interactions likely further increased when electron-electron interactions (neglected in the above analysis) are considered. Nonetheless, we find that the underlying self-energies are well described within the standard weak-coupling Migdal-Eliashberg framework. Fig.~\ref{fig:CoO}(e) shows the real and imaginary part of the self-energy ($\Sigma'$ and $\Sigma"$, respectively) extracted for the inner band along $\Gamma$-K (see Appendix; similar results were also obtained for the other states where the DFT provides a reasonable bare band description as shown in Supplementary Fig.~6). The full self-energy is well described by a simple two-oscillator Debye-model self-energy, with phonon modes at energies of $\omega_1=78\pm 5$~meV and $\omega_2=130\pm10$~meV (Fig.~\ref{fig:CoO}(e)). The softer mode appears likely to reflect a coupling to the Co-O $E_{u}$ or $A_{1g}$ optical phonon modes~\cite{homesPerfectSeparationIntraband2019b, chengRolePhononPhonon2017b, kumarFirstPrinciplesStudy2013a, takatsuRolesHighFrequencyOptical2007a} at this CoO$_2$-terminated surface. The higher mode appears too high for the known bulk phonon modes, but could reflect a harder phonon at the surface, a two-phonon loss process, or additionally also interband scattering between the surface and the bulk states, which cross close to this energy, and where additional increase in the linewidth appears to onset in our data shown in Fig.~\ref{fig:CoO}(d). The total coupling strength extracted from our self-energy fits results in $\lambda = 0.9\pm 0.1$ which is in excellent agreement with that determined directly from the measured Fermi velocity above. The electron-phonon interaction at the CoO$_2$-terminated surface therefore appears entirely conventional in nature. The enhanced coupling strength compared to the bulk can be attributed to the distinct atomic orbitals involved in the interaction and the significantly larger density of states at the Fermi level for these surface states as compared to the bulk.

\section{Polaron formation at the P\lowercase{d}-terminated surface}
\begin{figure*}
    \centering
    \includegraphics[width=\textwidth]{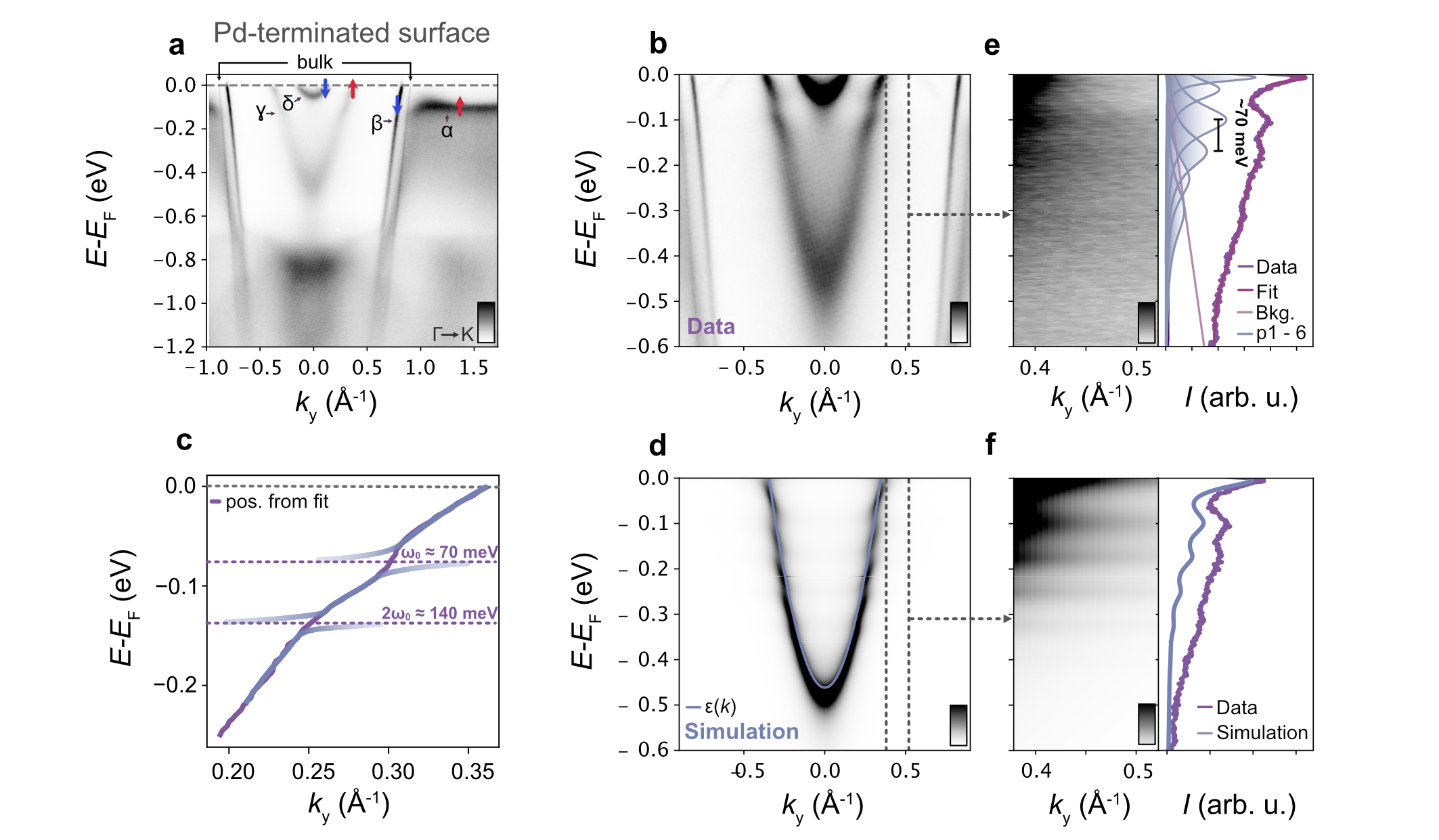}
    \caption{(a) Measured electronic structure of the Pd-terminated surface, along the $\Gamma$-K direction ($h\nu=90$~eV). This shows the feint but extremely sharp bulk state, and the $\alpha$-$\beta$ and $\gamma$-$\delta$ exchange-split surface state partners.  
    (b) Measurements of the $\gamma$ band ($h\nu=68$~eV), showing the formation of a ladder of kink-like features, which are clearly visible in fits to MDCs ((c), with the blue solid lines showing a guide to the eye), with strongly modulated spectral weight. 
    (d) Similar spectral signatures are visible in spectral function simulations for a Holstein polaron (see Appendix for the relevant parameters used). 
    (e,f) Notably, multiple replicas of dispersionless phonon-like branches are visible in the background spectra of both the measured (e) and calculated (f) spectral functions, extracted from the dashed regions shown in (b) and (d), respectively. The EDC integrated over this momentum region is shown, together with multi-peak fits which yield a sequence of modes separated by 70~meV.
    }
    \label{fig:Pd}
\end{figure*}
The situation is, however, rather different again for the Pd-terminated surface. We show an overview of the electronic structure of this surface from our $\mu$-ARPES measurements in Fig.~\ref{fig:Pd}(a). This electronic structure has previously been shown to reflect a Stoner instability to ferromagnetism at the Pd-terminated surface~\cite{mazzolaItinerantFerromagnetismPdterminated2018a}, with exchange-split $\alpha/\beta$ and $\gamma/\delta$ pairs of states clearly visible in Fig.~\ref{fig:Pd}(a). While this basic band structure has been observed previously \cite{mazzolaItinerantFerromagnetismPdterminated2018a,mazzolaTuneableElectronMagnon2022a}, here we resolve much sharper spectral features, in particular without additional spectral weight contributed from the CoO$_2$ surface states which dominate the measured signal for mixed surface terminations in conventional ARPES (see also our Fermi surface measurements in Supplementary Fig.~7).

From our high-resolution measurements, we find, in particular, that the electron-like $\gamma$-band exhibits rather striking signatures of self-energy modulations. Unlike the conventional kink structure observed at the CoO$_2$-terminated surface (Fig.~\ref{fig:CoO}), the spectral weight of the $\gamma$ band exhibits a pronounced peak-dip-peak-dip-... ladder-like structure, more indicative of the opening of a sequence of local band gaps in the spectral function. Such a behaviour indicates the scattering of electrons by multiple optical phonons, leading to an anti-crossing-like behaviour in the measured electronic structure~\cite{goodvinGreenFunctionHolstein2006,veenstraSpectralFunctionTour2011}. Consequently, the bands are flattened, with gaps opening in the dispersion at integer numbers of the characteristic phonon mode energy (see blue solid lines in Fig.~\ref{fig:Pd}(c)). This opening of new band gaps results in the steep kink-like features observed in the fitted peak position extracted from the MDCs shown in Fig.~\ref{fig:Pd}(c). Such a behaviour has been theoretically predicted for the formation of polarons within the Holstein model~\cite{holsteinStudiesPolaronMotion1959, holsteinStudiesPolaronMotion1959a, goodvinGreenFunctionHolstein2006, veenstraSpectralFunctionTour2011}, and was recently proposed to explain the unusual spectra measured in doped MoS$_2$, MoSe$_2$ and A$_{0.3}$MoO$_2$ [A=K, Rb]\cite{kangHolsteinPolaronValleydegenerate2018, jungHolsteinPolaronsRashbalike2024, kangBandselectiveHolsteinPolaron2021}. Indeed, very similar spectral signatures to the ones we observe here can be reproduced via momentum-average approximation calculations of the Holstein Hamiltonian performed with parameters relevant for the $\gamma$ band of the Pd-terminated surface here (Fig.~\ref{fig:Pd}(d), see Appendix for details).

\begin{figure*}
    \centering
    \includegraphics[width=\textwidth]{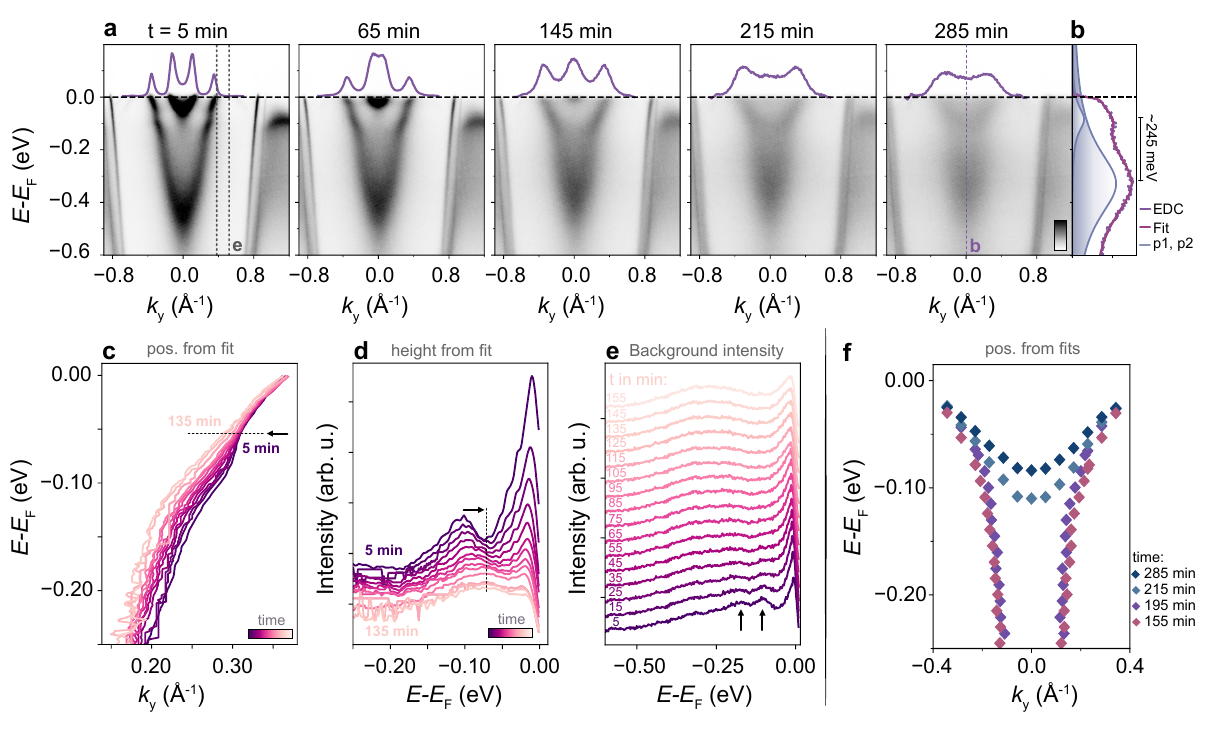}
    \caption{Adsorption-dependent evolution of the electron-boson coupling for Pd-terminated PdCoO$_2$. (a) Measured electronic structure of the Pd-terminated surface ($h\nu=68$~eV) as a function of time, $t$ [$t=0$ is the start of the measurement sequence, around 10 minutes after cleaving the sample, see Appendix]. An MDC at the Fermi level is shown for each measurement (purple line).
    (b) EDC at $\Gamma$ for the spectrum measured at $t=285$~min, with a two-peak fit as shown.
    (c,d). Peak position (c) and height (d) of the $\gamma$ band from fits to MDCs as a function of time from the start of the measurements until $t=135$~mins. 
    (e) Corresponding $t$-dependence of the background spectrum over the same time period, integrated in momentum over the dashed region shown in (a). 
    (f) Extracted dispersions from EDC and MDC fits to the dispersions for the later time points, showing the formation of a strongly renormalised quasiparticle band between $t=195$~min and $t=215$~min. 
    }
    \label{fig:Pd_time}
\end{figure*}

The calculations show the formation of sub-bands spaced by an energy equal to the relevant phonon mode energy $\omega_0$. Importantly, the mixing of the electron and phonon character should render the original near-dispersionless phonon replica branches visible in our photoemission measurements close to the dispersive band (see the guide to the eye in Fig.~\ref{fig:Pd}c). A key spectroscopic fingerprint here of the formation of polarons, therefore, would be a background spectrum comprising a ladder of flat states arising from such optical phonon branches, which are equally spaced by $\omega_0$, as shown in the magnified view of the simulated spectral function in Fig.~\ref{fig:Pd}f). We show in Fig.~\ref{fig:Pd}(e) that exactly such a background spectrum is obtained from our measured spectral function in Fig.~\ref{fig:Pd}(b). 

Our spectroscopic data thus hosts the key signatures expected from the formation of polarons. The local gaps that open are small, however, and the majority of the spectral weight remains in the dispersive branch, pointing to the formation of large polarons~\cite{holsteinStudiesPolaronMotion1959}. Such large polarons are well known to be stabilised in the intermediate carrier density and coupling regime~\cite{Perroni_PhysRevB.103.245130}, for example when charge carriers are doped into ionic semiconductors~\cite{dickeyPolaronEffectsCyclotronResonance1967, scottResonantRamanScattering1970, moserTunablePolaronicConduction2013a}. Indeed, such features are well studied in the bulk and at the interfaces of various transition-metal oxides~\cite{moserTunablePolaronicConduction2013a,  wangTailoringNatureStrength2016a, swartzPolaronicBehaviorWeakcoupling2018, reticcioliCompetingElectronicStates2022,chenObservationTwodimensionalLiquid2015,wangSurfaceChemistryPolarizable2022,  rileyCrossoverLatticePlasmonic2018, cancellieriPolaronicMetalState2016,yukawaPhonondressedTwodimensionalCarriers2016,ellingerSmallPolaronFormation2023}. However, in these systems, the polar electron-lattice interaction has been observed to be readily screened by the induced charge carriers. This typically leads to a crossover from a polaronic to a Fermi liquid regime with increasing carrier doping~\cite{verdiOriginCrossoverPolarons2017a, wangTailoringNatureStrength2016a, kangHolsteinPolaronValleydegenerate2018, Perroni_PhysRevB.103.245130}. Here, however, we are deep in the anti-adiabatic regime. A key question is therefore how such a polaronic coupling can persist, despite the extreme metallicity of this surface.

From fitting our observed background spectra shown in Fig.~\ref{fig:Pd}(e), we obtain a  characteristic mode energy, $\omega_0 \approx 70$~meV. This agrees well with literature values for the longitudinal optical (LO) $A_{2u}$ phonon mode, consisting of an out-of-plane Pd-O vibration, and which has been shown to couple with the bulk $ab$-plane carriers in PdCoO$_2$~\cite{homesPerfectSeparationIntraband2019b, seoInteractionInplaneDrude2023,yaoOriginHighElectrical2024}. The mobile electrons of the Pd-terminated surface are, however, highly two dimensional, confined to the surface plane. They are not, therefore, able to effectively screen the electric field resulting from the out-of-plane polar phonon mode. We thus attribute the strong electron-phonon coupling of the $\gamma$ band here to a mode-selective suppressed screening, despite the highly-metallic nature of the surface. This, in fact, suggests that coupling to out-of-plane polar modes could provide a rather general route to the formation of polaronic states at 2D surfaces and interfaces.
 
\section{Surface adsorption-controlled polaronic coupling}
Interestingly, we find that spectra of the quality and form shown in Fig.~\ref{fig:Pd}(a,b) are only found within the first few minutes of cleaving the sample. We show in Fig.~\ref{fig:Pd_time} the evolution of the measured electronic structure of the Pd-terminated surface over a time period of 290~mins. Already in the raw data, significant changes in the electronic surface structure can be observed, with an increase in observed linewidth, a depopulation of the $\delta$ band, and further subtle band-dependent energy shifts (see Supplementary Fig.~8 for a Luttinger analysis) that overall point to a small but resolvable increase in electron doping at the surface. These changes indicate that residual gases from the vacuum are becoming adsorbed on the sample surface with time. The short times in which these changes occur point to a highly reactive surface, in particular in comparison to the CoO$_2$-terminated surfaces of the same compound (Fig.~\ref{fig:CoO}), which we find are rather stable with time in the same measurement system.

Intriguingly, we find that the broad-scale band structure changes of the Pd-terminated surface are accompanied by marked variations in the polaronic coupling discussed above. Indeed, it is known that surface polarons can be strongly influenced by adsorption~\cite{reticcioliInterplayAdsorbatesPolarons2019, chengCOAdsorbatePromotes2022}, and we show in Figs.~\ref{fig:Pd_time}(c-e, respectively) how the MDC peak positions and peak heights as well as the extracted `background' spectra all evolve as a function of time within the initial period following sample cleavage. The extracted quasiparticle dispersions (Fig.~\ref{fig:Pd_time}(c)) reveal that the multi-kink/step-like structure described above is gradually smeared out with increasing time. This is accompanied by a continuous filling of the dips in spectral weight at the same energies (Fig.~\ref{fig:Pd_time}(d)). This points to a closing of the local band gaps between the different sub-bands of the renormalised dispersion. Consistent with this, we find a gradual suppression of the spectral weight of the flat-band phonon-like replica features in the background spectra (Fig.~\ref{fig:Pd_time}(e)). Together, all of these spectral signatures indicate a strong decrease in the electron-phonon coupling strength of the surface electron gas to the Pd-O phonon mode, pointing to a screening of the electron-phonon coupling by the surface adsorption. 

The weakening of this polaronic coupling, however, is not the only change in the many-body spectrum which results from this adsorption. The background spectra in Fig.~\ref{fig:Pd_time}(e) show how a broad additional peak centred at an energy of $\approx\!270$~meV develops with time and becomes particularly pronounced towards the later time stages of our measurements. Simultaneously, we find that the dispersion of the $\gamma$-band (Fig.~\ref{fig:Pd_time}(a,f)) evolves from having an occupied bandwidth on the order of 500~meV, to having a strongly renormalised bandwidth of only $\approx 90$~meV, with a strong satellite feature located at a binding energy of $\approx250$~meV below the quasiparticle band bottom (Fig.~\ref{fig:Pd_time}b and f). This single broad peak-dip-hump structure bears the spectral hallmarks of a quasiparticle band strongly coupled to a high-energy bosonic mode, in line with the more typical polaron signatures observed by ARPES in, e.g., 2D electron gasses in SrTiO$_3$ and TiO$_2$~\cite{wangTailoringNatureStrength2016a, chenObservationTwodimensionalLiquid2015,moserTunablePolaronicConduction2013a}, but here with an even larger characteristic mode energy. 

\section{Discussion and Perspectives}
The high energy polaron mode, which develops at the later time stages, has previously been attributed by some of us to the coupling to a magnon~\cite{mazzolaTuneableElectronMagnon2022a}. While we cannot exclude that magnon modes still contribute to the self-energy at these energy scales, our new observations motivate a reconsideration of the dominant effects in the measured spectra. In particular, the systematic observation of how the coupling to this high energy mode increases with surface adsorption suggests that a hard vibrational mode of the surface adsorbate may provide a better explanation for the spectroscopic signatures observed. Pd is known to be highly reactive to the adsorption of H~\cite{liHydrogenStoragePd2014,alefeldHydrogenMetals1978, kawaeSuperconductivityPalladiumHydride2020}, which is typically the primary residual species in an ultra-high vacuum chamber. This therefore presents a leading candidate for the adsorbed species here: for the analogous system of H/Pt(111), hydrogen adsorption is known to give rise to new phonon modes with energies up to $\approx275$~meV~\cite{hongFirstprinciplesCalculationsPhonon2005}, comparable with the energy scales observed here. Previous studies of bulk PdCoO$_2$, meanwhile, have shown promise for use in electrocatalysts with high hydrogen evolution reaction activity~\cite{liSituModificationDelafossiteType2019, podjaskiRationalStrainEngineering2019} linked to the formation of Pd nanoclusters at the surface that exhibit high hydrogen sorption/desorption reversibility. Together, our findings thus suggest that atomic or molecular adsorption could provide a controllable route to tune the polaronic coupling in PdCoO$_2$, in turn potentially also influencing its catalytic properties~\cite{reticcioliInterplayAdsorbatesPolarons2019, franchiniPolaronsMaterials2021, renRecentProgressesPolarons2024}. Here, we find spectroscopically that it drives a crossover between two markedly-distinct polaronic regimes. 

We can gain these insights because of the exceptional data quality obtainable from this model system, and due to the spatial selectivity but high spectral resolution offered by modern capillary-based $\mu$-ARPES. Here, these experiments allow not only distinguishing the distinct electronic structures of different surface terminations, but also unveil their disparate quasiparticle dynamics and surface reactivity. Despite the fact that both the CoO$_2$- and Pd-terminated surfaces support a strong electron-phonon coupling, only the latter yields polaronic signatures, notwithstanding its more itinerant nature. This points to a key role of the symmetry of the relevant phonon modes involved in the coupling, here leading to a poorly-screened mode that can selectively couple to the electronic states of the Pd-terminated surface. While the itinerant carriers in the 2D surface plane are ineffective at screening this mode, we have found that surface adsorption provides a powerful route to tune this coupling. Together, this suggests routes to the deterministic control of electron-phonon interactions at material interfaces, for example via selective geometries in superlattice configurations, or by interfacing solid state and molecular layers, with potential implications for tuning e.g. thermoelectric properties or developing optimised routes to stabilise interface superconductivity. 

\section*{Acknowledgements}
We thank Chun Lin, Peter Littlewood, Federico Mazzola, Bipul Pandey, Valentina de Renzi, Michele Reticcioli, Veronika Sunko, Andrea Tonelli, and Peter Wahl for useful discussions. 
We gratefully acknowledge support from the European Research Council (through the QUESTDO project, 714193), the Leverhulme Trust (Grant No.~RPG-2023-256) and the UK Engineering and Physical Sciences Research Council (Grant No.~EP/T02108X/1).
We thank Diamond Light Source for access to the I05 beamline (Proposal Nos.~SI28445 and SI30125), which contributed to the results presented here.
For the purpose of open access, the authors have applied a Creative Commons Attribution (CC BY) licence to any Author Accepted Manuscript version arising. 

\section*{Data Availability Statement}
The research data that support the findings of this article will be made openly available before publication. 

\appendix
\section{Methods}

\subsection{Angle-resolved photoemission} Single crystal samples measured in this study have been grown by flux and vapor transport in a quartz tube. All samples have been cleaved $in$ $situ$ at a base pressure of $\approx 1\times10^{-10}$~mbar to obtain a clean surface. Angle-resolved photoemission measurements were performed at the nano-ARPES branch of the I05 endstation at Diamond Light Source, UK, using an elliptical capillary focusing optic leading to a spot size on the sample of $\sim$ 4 $\mu $m. The photoemitted electrons were detected using a Scienta Omicron DA30 electron analyser. A piezoelectrically-driven 5-axis sample manipulator enabled fast spatial mapping of the sample. All measurements were performed at the base temperature of $T\approx$ 35 K using linear horizontal ($p$-) polarised light with photon energies in the range of $h\nu = 54 - 170$~eV. For the time-dependent evolution of the electronic structure shown in Fig.~\ref{fig:Pd_time} the sample was rapidly aligned on a Pd terminating surface immediately after cleaving. The time between cleaving and starting the measurement series was kept to a minimum. To track the time dependence of the measured dispersions, we then took a series of measurements with an acquisition time of 1~min each. $t=0$ is defined by the start of the first measurement of the series, approximately 10~min after cleaving the sample. To increase statistics, ten spectra were binned together, which reflects an appropriate compromise between statistics and the timescale on which non-negligible changes in the electronic structure occur. \\

\

\subsection{Density functional theory} 
Electronic structures were calculated using a local-orbital minimum basis method implemented in the full-potential FPLO code~\cite{Koepernik1999,Opahle1999} version fplo22.00-62 (http://www.fplo.de). The calculations were based on the local density approximation (LDA) with Perdew-Wang-92 exchange-correlation function~\cite{Perdew1992} and on the generalized gradient approximation (GGA) with the Perdew-Burke-Ernzerhof exchange-correlation function~\cite{Perdew1996}. The spin-orbit coupling effect was treated non-perturbatively by solving the full Kohn-Sham-Dirac equation~\cite{Eschrig2004} in a single-step calculation. The Brillouin zone was sampled with a well converged $k$-mesh of 1600 $k$-points ($20\times20\times4$ mesh, 165 points in the irreducible wedge of the Brillouin zone). For surface electronic structures we constructed periodic arrangements of Pd and O terminated slab structures including nine Co layers, using the bulk crystal structure of PdCoO$_2$ with a thick vacuum of 15~\AA{} between adjacent slabs. For the construction, the bulk structure in the rhombohedral space group $R\bar{3}m$ was used, with the experimental lattice parameters of $a$ = 6.1359~\AA{} and a rhombohedral angle of 26.67 degree at room temperature. To facilitate the convergence of the self-consistent calculations, we formed centrosymmetric cells by applying a combination of $C_2$ rotation and mirror symmetry about the central Co layer of the supercell. We then constructed distinct slab structures with Pd and CoO$_2$ surface terminations. To assure the convergenc with respect to the slab thickness we compared the density of states (DOS) of the central layer with the corresponding bulk DOS, obtaining very similar results.

\

\subsection{Self-energy extraction} We extract the peak positions and linewidths from fits to MDCs extracted form our experimental data. To convert these into self-energies, we utilise a bare band derived from our DFT surface supercell calculations. Supercell calculations are challenging for polar materials, with off-stoichiometry of the resulting supercell leading to difficulties in assigning absolute energy positions. We therefore shift the bands by $92\pm 5$~meV to match $k_\text{F}$ with the experimental data, and scale the bandwidths by a global factor of 1.73 in order to match the experimentally determined position of the surface states at the K-point. This approach thus effectively removes the effect of electron-electron interactions, providing a bare band for use in the extraction of the electron-phonon self-energy. The real part of the self-energy $\Sigma'$ has then been extracted as the energetic difference between the measured dispersion and the so determined bare band while the imaginary part $\Sigma"$ is given by FWHM$=2\vert \Sigma"(E-E_\text{F})/v^{DFT}\vert$, where $v^{DFT}$ describes the velocity of the bare band extracted from our calculations. Due to the strong variations of $v^{DFT}$ for different binding energies (see supplementary Fig. 4), we used a varying velocity $v^{DFT}(E-E_\text{F})$. The so-extracted imaginary part of the self-energy was fit to a model self-energy derived within the Midgal-Eliashberg theory using the Debye model. This is well described in Refs.~\cite{hofmannElectronPhononCoupling2009c, hellsingElectronphononCouplingMetal2002a, kirkegaardSelfenergyDeterminationElectron2005c}. The corresponding real part was determined via Kramer-Kronig transformation.

\

\subsection{Spectral function simulations} 
The model used in this work is based on the momentum average (MA) approximation which, as discussed in detail in Ref.~\cite{goodvinGreenFunctionHolstein2006,veenstraSpectralFunctionTour2011}, yields a spectral function expanded in powers of the coupling strength $g^2$ running opver all energies ($\omega$) as:
\begin{equation}
	\begin{split}
	\Sigma_{MA}(\omega)=&g^2 	\bar{g}_0(\omega-\omega_0) +g^4 [2\bar{g}_0^2(\omega-\omega_0)\bar{g}_0(\omega-2\omega_0)]\\
	&+ g^6[4\bar{g}_0^3(\omega-\omega_0)\bar{g}_0^2(\omega-2\omega_0)+6\bar{g}_0^2(\omega-\omega_0)\\
	&\times \bar{g}_0^2(\omega-2\omega_0)\bar{g}_0(\omega-3\omega_0)]+O(g^8),
	\end{split}
\end{equation}
where $\omega_0$ describes the critical mode energy. This has been used to model the spectral function of a single band derived by a simplified 2D triangular lattice tight binding model defined by the hopping parameter $t = 0.18$~eV  in Fig.~\ref{fig:Pd}(d,f) using a total coupling strength of $\lambda = g^2/(4t\omega_0)= 1.6$.

\bibliography{PdCoO2_SE.bib}

\end{document}


\title{Supplementary Information: Dichotomy of electron-phonon interactions in the delafossite PdCoO$_2$: From weak bulk to polaronic surface coupling}
\author{Gesa-R. Siemann}
\affiliation{SUPA, School of Physics and Astronomy, University of St Andrews, St Andrews KY16 9SS, UK}

\author{Philip A. E. Murgatroyd}
\affiliation{SUPA, School of Physics and Astronomy, University of St Andrews, St Andrews KY16 9SS, UK}

\author{Tommaso Antonelli}
\affiliation{SUPA, School of Physics and Astronomy, University of St Andrews, St Andrews KY16 9SS, UK}

\author{Edgar Abarca Morales}
\affiliation{SUPA, School of Physics and Astronomy, University of St Andrews, St Andrews KY16 9SS, UK}
\affiliation{Max Planck Institute for Chemical Physics of Solids, 01187 Dresden, Germany}

\author{Seunghyun Khim}
\affiliation{Max Planck Institute for Chemical Physics of Solids, 01187 Dresden, Germany}

\author{Helge Rosner}
\affiliation{Max Planck Institute for Chemical Physics of Solids, 01187 Dresden, Germany}

\author{Matthew D. Watson}
\affiliation{Diamond Light Source, Harwell Science and Innovation Campus, Didcot OX11 0DE, UK}

\author{Andrew P. Mackenzie}
\affiliation{Max Planck Institute for Chemical Physics of Solids, 01187 Dresden, Germany}
\affiliation{SUPA, School of Physics and Astronomy, University of St Andrews, St Andrews KY16 9SS, UK}

\author{Phil D. C. King}
\email{pdk6@st-andrews.ac.uk}
\affiliation{SUPA, School of Physics and Astronomy, University of St Andrews, St Andrews KY16 9SS, UK}

\date{\today}

\maketitle

\begin{figure*}
    \centering
    \includegraphics[width=\textwidth]{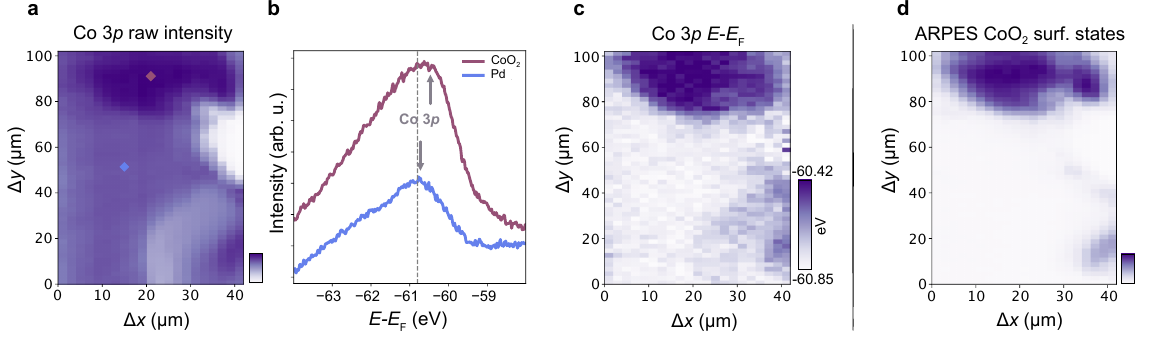}
    \caption{{Spatially resolved core-level photoemission of PdCoO$_2$.} 
    (a) Spatial variation of the Co 3$p$ intensity across the sample probed using a photon energy of 170 eV. (b) Extracted core level data from the map shown in (a) at positions indicated by the diamond markers. A clear difference in intensity as well as a subtle shift of the core level's binding energy can be observed for different positions on the sample surface. (c) The binding energy ($E-E_F$) of the Co 3$p$ core level across the probed sample region extracted from fits to the data. The decreased the binding energy of the core level component correlates well with those areas of the sample where the surface states of a CoO$_2$-terminated surface can be observed as shown in (d).
    }
    \label{supp:spatial}
\end{figure*}

\begin{figure*}
    \centering
    \includegraphics[width=\textwidth]{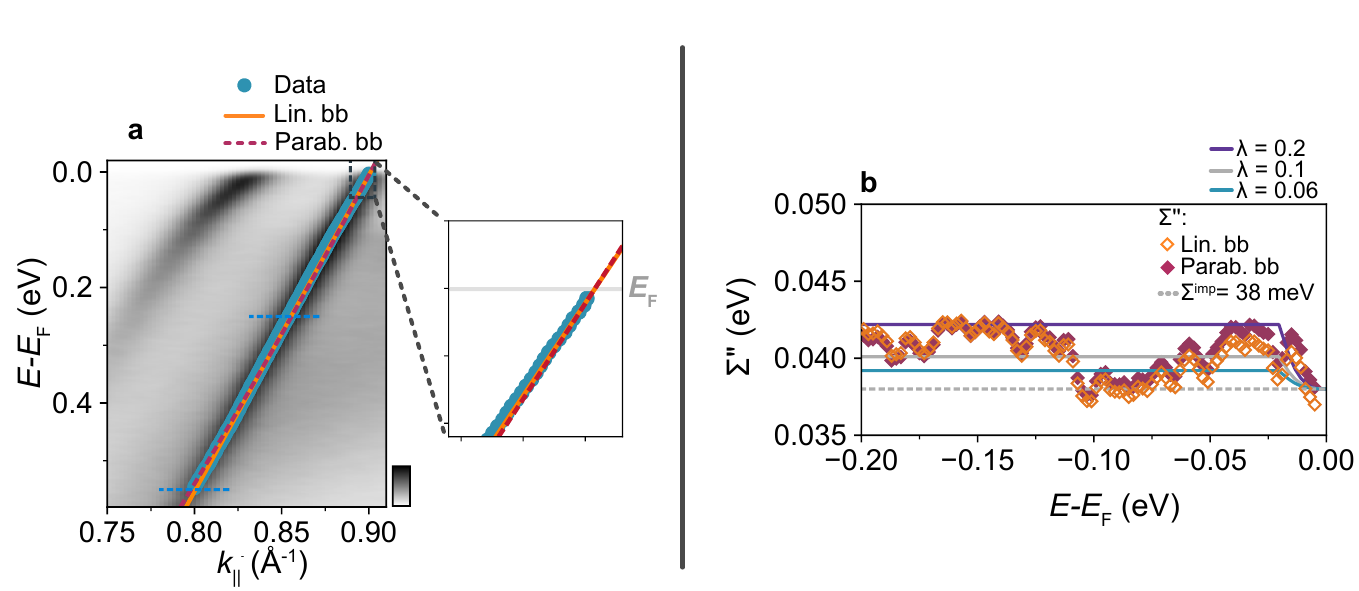}
    \caption{{Electron-phonon coupling of the bulk state. } 
(a) Magnified view of the ARPES spectra of the Pd terminated surface shown in Fig.~2a of the main paper. The peak positions extracted from MDC fits to the bulk band are shown by the blue markers. The linear (orange solid line) and parabolic (red dashed line) bare band (bb) approximations are shown. These are obtained from fits to the data at higher binding energies (between the blue dashed lines), with the Fermi wavevector pinned to the experimentally determined $k_\text{F}$. A further magnification is shown in the inset, which indicates the decreased slope of fits to the data in the vicinity of the Fermi level as compared to the bare bands. (b) Imaginary part of the self-energy $\Sigma''$ extracted from the linewidth from fits to MDCs and converted to the self energy using  a fixed (orange markers) and an energy dependent bare band velocity $v(E-E_{\text{F}})$ (red markers) as pertains to the linear and parabolic bare band approximations, respectively. Debye model self energy simulations are shown for different values of the electron-phonon coupling constant, $\lambda$, for a phonon mode of $\hbar \omega = 20$~meV (coloured lines), as well as a constant impurity term of 38~meV.
    }
    \label{supp:bulkbb}
\end{figure*}

\begin{figure*}
    \centering
    \includegraphics[width=\textwidth]{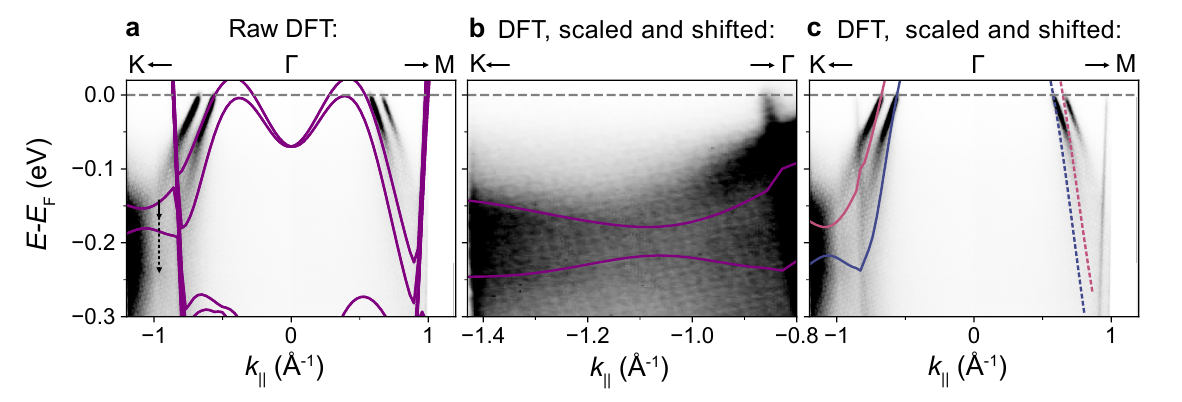}
    \caption{{DFT calcualations of the CoO$_2$ terminated surface of PdCoO$_2$. } 
    (a) As-calculated DFT dispersions plotted on top of the ARPES data, indicating a strong underestimatation of the band width and overestimataion of the Fermi level of the calculations. To correct for this, and still allow the $k$-dependent form of the dispersion to be used as a good appoximation to the bare band dispersion, we shift the dispersion by $92\pm 5$~meV and scale it about the Fermi level by a factor of 1.73. (b,c) This (b) matches the calculations to the data at the band bottom close to the K - point (h$\nu = 76$~eV), effectively removing any impact of electron-electron interactions, and (c) attempts to best reproduce the Fermi crossings.
    }
    \label{supp:DFT}
\end{figure*}

\begin{figure*}
    \centering
    \includegraphics[scale=0.5]{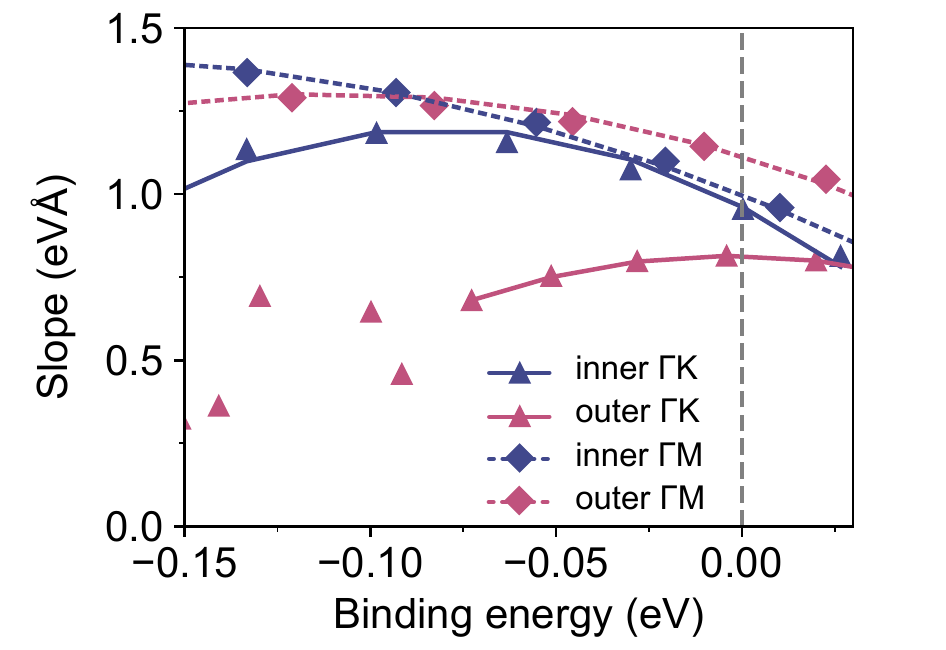}
    \caption{{Non-linear bare band dispersions.} Extracted gradient of the DFT calculations shown in Fig.~\ref{supp:DFT} along the $\Gamma$-M (diamond markers) and $\Gamma$-K (triangle markers) direction. A parabolic fit to the extracted band slopes (dashed and solid lines) highlights the strongly non-linear nature of the underlying band dispersions at and in the vicinity of the Fermi level, and thus highlights that it would be an inaccurate approach to use a linear bare band approximation.
    }
    \label{supp:Luttinger}
\end{figure*}

\begin{figure*}
    \centering
    \includegraphics[width=\textwidth]{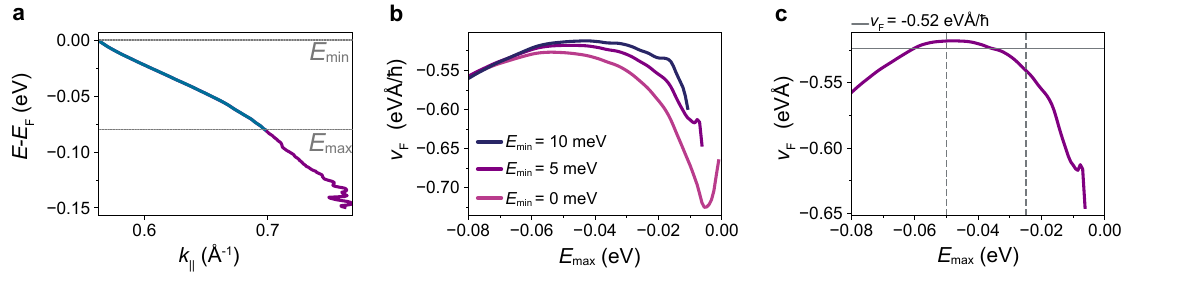}
    \caption{{Illustration of the method used to extract the Fermi velocity of the surface states.} (a) The extracted peak positions from MDC fits to the inner band along the $\Gamma$-K direction (purple markers) highlighting a low energy cut-off ($E_{\text{min}}$) and a high energy cut-off ($E_{\text{max}}$) used to determine the linear fit to the data to extract $v_\text{F}$. (b) The resulting $v_\text{F}$ as a function of the high energy cut-off for different values of the low energy cut-off, illustrating the effect of both finite energy resolution and space charge. These can lead to an overestimation of $v_\text{F}$ for too small values of $E_{\text{min}}$, but an underestimation for value that are too high. Given the extended range of stability in extracted $v_\text{F}$, and considering the relevant energy resolutions here, we consider $E_{min}=5$~meV provides an effective compromise. (c) The experimental $v_\text{F} = -0.52\pm 0.02$~eV\AA~(solid line) has then been extracted as the average value determined by fitting the data with $E_{\text{max}}$ varied between 25 and 50~meV (dashed lines).
    }
    \label{supp:spatial}
\end{figure*}

\begin{figure*}
    \centering
    \includegraphics[scale=0.8]{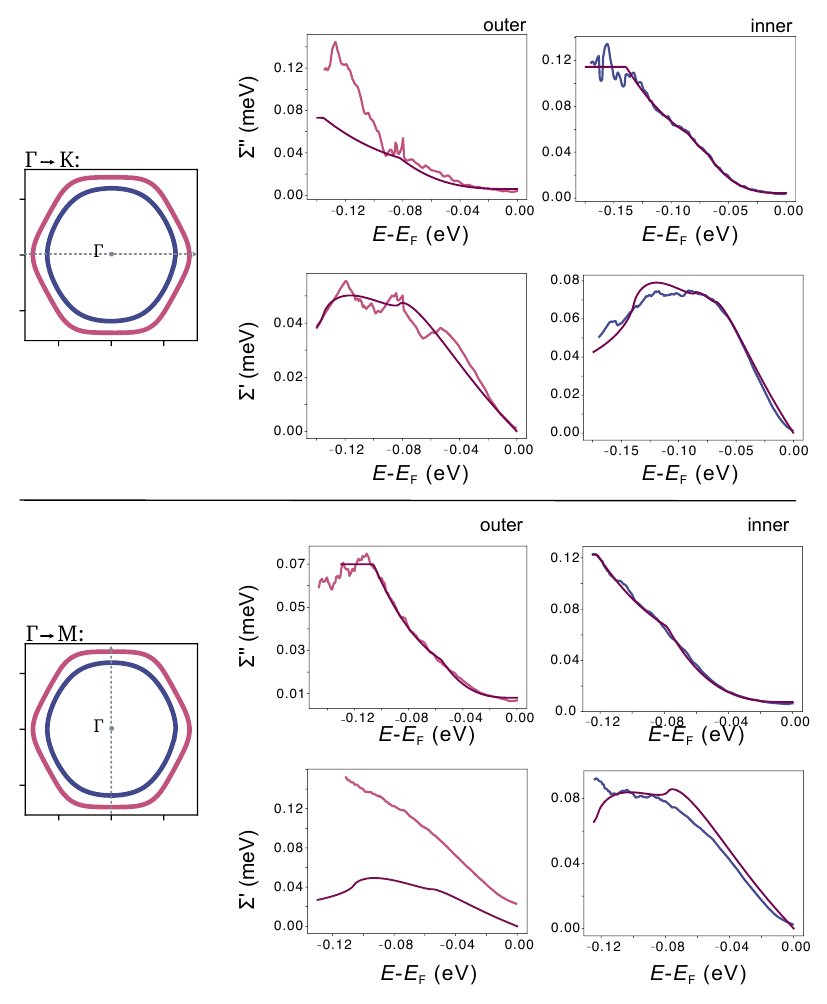}
    \caption{{Fits to the experimentally determined self energy.} The real $\Sigma'$ and imaginary $\Sigma"$ parts of the self energy have been extracted using the surface projected DFT calculations shown in Supplementary Fig.~\ref{supp:DFT} as bare bands. The imaginary part of the self energy has then been fitted using the Midgal-Eliashbaerg approach and a Debye model. The real part of the self energy can then be extracted via a Kramers-Kronig transformation of the fit. This approach however fails for the outer bands along the $\Gamma$-M direction, due to the mismatch of the data and the DFT calculations as discussed in the main manuscript of this paper.
    }
    \label{supp:Luttinger}
\end{figure*}

\begin{figure*}
    \centering
    \includegraphics[scale=0.7]{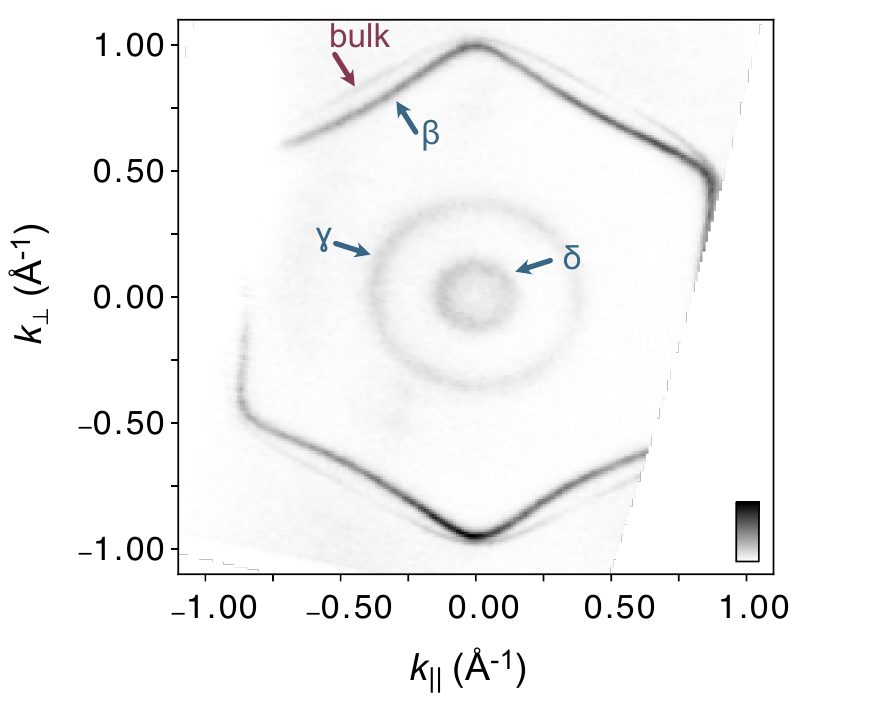}
    \caption{{Fermi surface.} Fermi surface of a Pd-terminated surface of PdCoO$_2$ measured using a photon energy of 90~eV and LH polarised light. The surface states that cross the Fermi level, and thus contribute to the Fermi-surface, are labeled $\beta$, $\gamma$, $\delta$. 
    }
    \label{supp:spatial}
\end{figure*}

\begin{figure*}
    \centering
    \includegraphics[width=\textwidth]{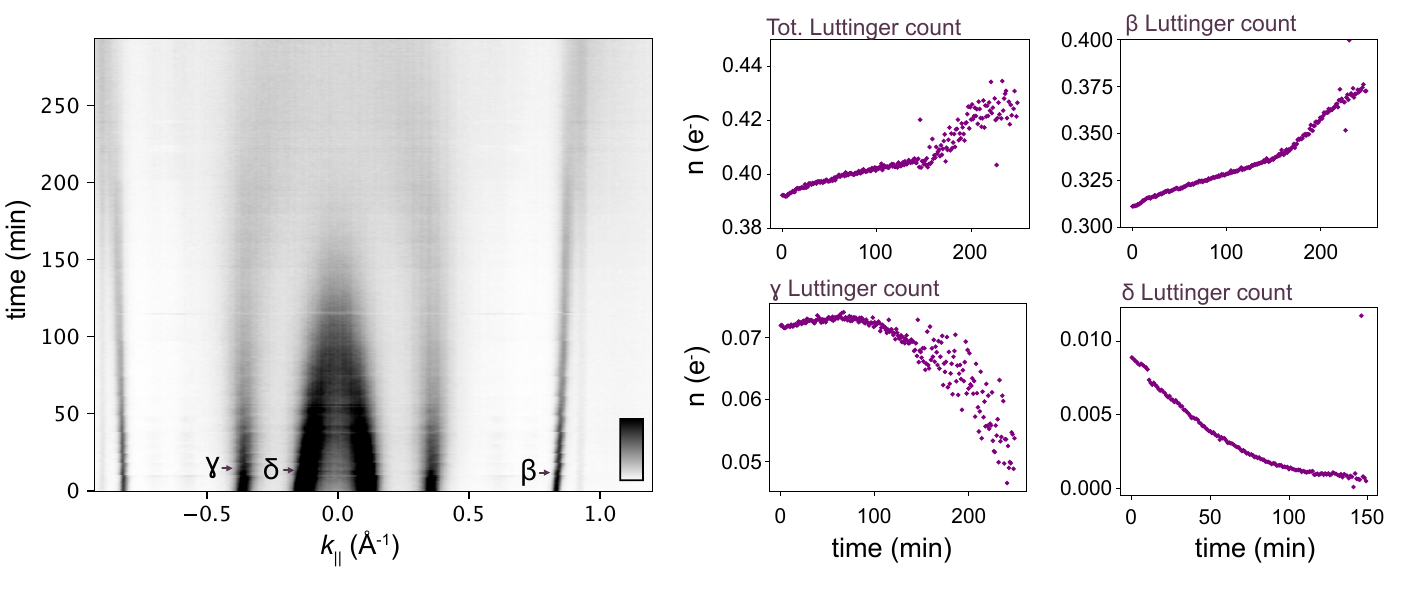}
    \caption{{Adsorption-dependent band shifts.} Changes in the Fermi wave vector along the $\Gamma$-K direction as a function of time (left). (Right) The extracted total Luttinger count and that for the different surface states ($\beta$, $\gamma$ and $\delta$). While the minority state at the $\Gamma$-point (labelled $\delta$) shift upwards and thus decreases $k_\text{F}$ and associated carrier count, the main contribution to the total Luttinger count is given by the $\beta$-state, whose $k_\mathrm{F}$ slightly increases with time. As a consequence, the overall carrier count slightly increases with time suggesting a steady electron doping of the surface.} 
    \label{supp:Luttinger}
\end{figure*}

\bibliographystyle{}
\bibliography{references}